\documentclass[%
 aip,
 amsmath,amssymb,
 reprint,%
]{revtex4-1}

\usepackage{graphicx}
\usepackage{dcolumn}
\usepackage{bm}
\usepackage{siunitx}
\usepackage[utf8]{inputenc}
\usepackage[T1]{fontenc}
\usepackage{mathptmx}
\usepackage{etoolbox}
\usepackage{comment}
\usepackage{amsmath}
\usepackage{setspace}
\usepackage{array}
\usepackage{booktabs}
\usepackage{amsbsy} 
\usepackage{amsthm} 
\usepackage{mathptmx}   
\usepackage{float}
\usepackage{soul}
\usepackage[colorlinks=true,linkcolor=blue,citecolor=blue,allcolors=blue]{hyperref}

\usepackage[dvipsnames,svgnames,x11names]{xcolor}

\makeatletter
\def\@email#1#2{%
 \endgroup
 \patchcmd{\titleblock@produce}
  {\frontmatter@RRAPformat}
  {\frontmatter@RRAPformat{\produce@RRAP{*#1\href{mailto:#2}{#2}}}\frontmatter@RRAPformat}
  {}{}
}%
\makeatother
\begin{document}
\preprint{AIP/123-QED}

\title{Acoustofluidic Suppression of Rayleigh–Taylor Instability and Fluid Mixing: Stabilization of Stratified Fluids in a Minichannel}

\author{Venkatesh Seenuvasan Revathi$^{\dagger}$}
\author{Jeyapradhap Thirisangu$^{\dagger}$}
\author{Karthick Subramani*}
 \email{karthick@iiitdm.ac.in}
\affiliation{$^1$Department of Mechanical Engineering, Indian Institute of Information Technology, Design and Manufacturing Kancheepuram, Chennai-600127, India.}

\date{\today}

\begin{abstract}

Rayleigh-Taylor Instability (RTI) typically arises when a dense fluid is superimposed on a lighter fluid, where the destabilizing gravitational force acting on miscible fluids drives chaotic mixing. We theoretically present an acoustofluidic method utilizing standing bulk acoustic waves (BAW) to counteract RTI and suppress the mixing of fluids. To successfully achieve this suppression, we demonstrate that two concurrent conditions are to be satisfied: the acoustic energy density ($E_{ac}$) of the standing waves must exceed its critical threshold ($E_{cr}$), and the orientation of the acoustic waves must be perpendicular to the fluid-fluid interface. This acoustofluidic mechanism reduces the mixing index (MI) by up to an order of magnitude compared to the mixing induced solely by gravity. By analyzing the interplay between acoustic and gravitational forces, this study provides a comprehensive understanding of acoustically modulated mixing dynamics in minichannels.

\end{abstract}

\maketitle

\section{{\textbf{INTRODUCTION}}}

The evolution of acoustofluidics encompasses a rich and diverse history, spanning from early fundamental investigations to modern microfluidic applications. The pioneering works of Faraday,\cite{Michael1831Dec} Kundt,\cite{Kundt1874Jan} Lord Rayleigh,\cite{Rayleigh1896SoundVol2} Hertz and Mende,\cite{hertz_mende_1939} and Westervelt \cite{Westervelt1953Jan} established a deep understanding of core phenomena such as acoustic radiation force, acoustic streaming, and the acoustic fountain effect. These foundational studies have paved the way for using acoustic fields as a versatile mechanism for applications ranging from bioparticle sorting and droplet manipulation to precise fluid control.\cite{FriendYeo2011} Leveraging these classical principles, the present work introduces an acoustofluidic approach to inhibit Rayleigh-Taylor instability (RTI) and suppress the mixing of fluids.

Lord Rayleigh\cite{Rayleigh1883} theoretically proved that the equilibrium of a dense fluid superimposed on a lighter fluid is inherently unstable under gravitational acceleration, thereby establishing the governing stability criterion for stratified fluids. Subsequently, G.I. Taylor\cite{Taylor1950} generalized this framework, demonstrating that instability arises whenever the acceleration is directed from the denser fluid toward the lighter medium, irrespective of the force’s origin (e.g., gravity, electric or magnetic fields). This prediction received its first visual confirmation from Lewis,\cite{Lewis1950InstabilityLiquidSurfacesII} whose experiments demonstrated that first-order theory predicts initial wave growth but fails at the onset of non-linearity, marked by the emergence of large round-ended columns of air. Subsequent theoretical studies by Bellman and Pennington,\cite{BellmanPennington1954} Chandrasekhar,\cite{Chandrasekhar1961} and Duff et al.\cite{DuffHarlowHirt1962} elucidated the stabilizing roles of surface tension, viscosity, and diffusion. The pioneering numerical simulations by Daly\cite{Daly1967} captured the complete temporal evolution of RTI, illustrating the Kelvin-Helmholtz instabilities that drive the formation of the classic mushroom-shaped morphology. The suppression of the Rayleigh-Taylor instability (RTI) has been explored through a diverse range of stabilization strategies, including intrinsic fluid properties,\cite{TroyonGruber1971} temperature gradients,\cite{Burgess2001suppression, Alexeev2007Marangoni} geometric and kinematic constraints,\cite{SenStorer1997flowcurvature, Alqatari2020Nov} dynamic mechanical strategies,\cite{Wolf1969dynamic, Wolf1970dynamic, Carnevale2002Apr, Xie2017Feb, GruberTroyon1971} and external electric\cite{Barannyk2012Feb, Cimpeanu2014control} and magnetic fields.\cite{Rannacher2007Jan} Suppression of Rayleigh-Taylor instability using an external acoustic body force has not yet been demonstrated.

Recent experimental investigations in microfluidic systems demonstrate that acoustic radiation force can relocate laminated liquid streams within a microchannel when subjected to a resonant acoustic field, and can also counteract the gravitationally induced inclination of the liquid–liquid interfaces.\cite{DeshmukhBrzozkaLaurellAugustsson2014} The theoretical framework for the acoustic body force driving this fluid relocation was initially established by Karlsen et al.\cite{KarlsenAugustssonBruus2016} using the mean Eulerian pressure hypothesis. Subsequently, Rajendran et al.\cite{Rajendran2022} provided a more rigorous first-principles derivation, offering greater theoretical clarity without relying on such assumptions.
Furthermore, Rajendran et al.\cite{Rajendran2023} established the stability criteria for inhomogeneous miscible and immiscible fluids under acoustic fields. Their theoretical model demonstrated excellent agreement with prior experimental observations by Deshmukh et al.\cite{DeshmukhBrzozkaLaurellAugustsson2014} and Hemachandran et al.\cite{HemachandranKarthickLaurellSen2019} Utilizing the above theoretical framework, Thirisangu et al.~\cite{Thirisangu2025UnifiedDroplets} demonstrated the manipulation of droplets, including migration, deformation, and splitting, both experimentally and theoretically. Employing the acoustic relocation phenomenon, Pothuri et al.\cite{Pothuri2019Dec} enhanced fluid mixing using sound waves via an alternating multinode excitation strategy in a microchannel. In their work, gravitational effects were neglected, as acoustic forces typically dominate in the microfluidic regime. However, in mini-scale fluidic systems, gravity-driven mixing and instabilities become significantly prominent. The interplay between the acoustic body force and gravity acting on miscible fluids within minichannels can yield a rich and complex dynamical landscape that remains largely unexplored.

In this theoretical study, we present an acoustofluidic approach that utilizes the acoustic body force induced by standing bulk acoustic waves (BAW) to simultaneously inhibit Rayleigh-Taylor Instability (RTI) and suppress the mixing of fluids in minichannels. We investigate two distinct Configurations comprising horizontally and vertically stratified fluid domains. Our results establish the existence of the critical acoustic energy density ($E_{cr}$), above which fluid mixing is effectively suppressed and the RTI is completely inhibited. Counterintuitively, when the acoustic energy density is below its critical threshold, the applied acoustic field not only fails to stabilize the system but enhances the mixing rate compared to the baseline mixing condition ($E_{ac}=0$) driven solely by gravity.

\section{{\textbf{Physics of the problem}}}

\begin{figure}[h]
  \center
    \includegraphics[width= 1\linewidth]{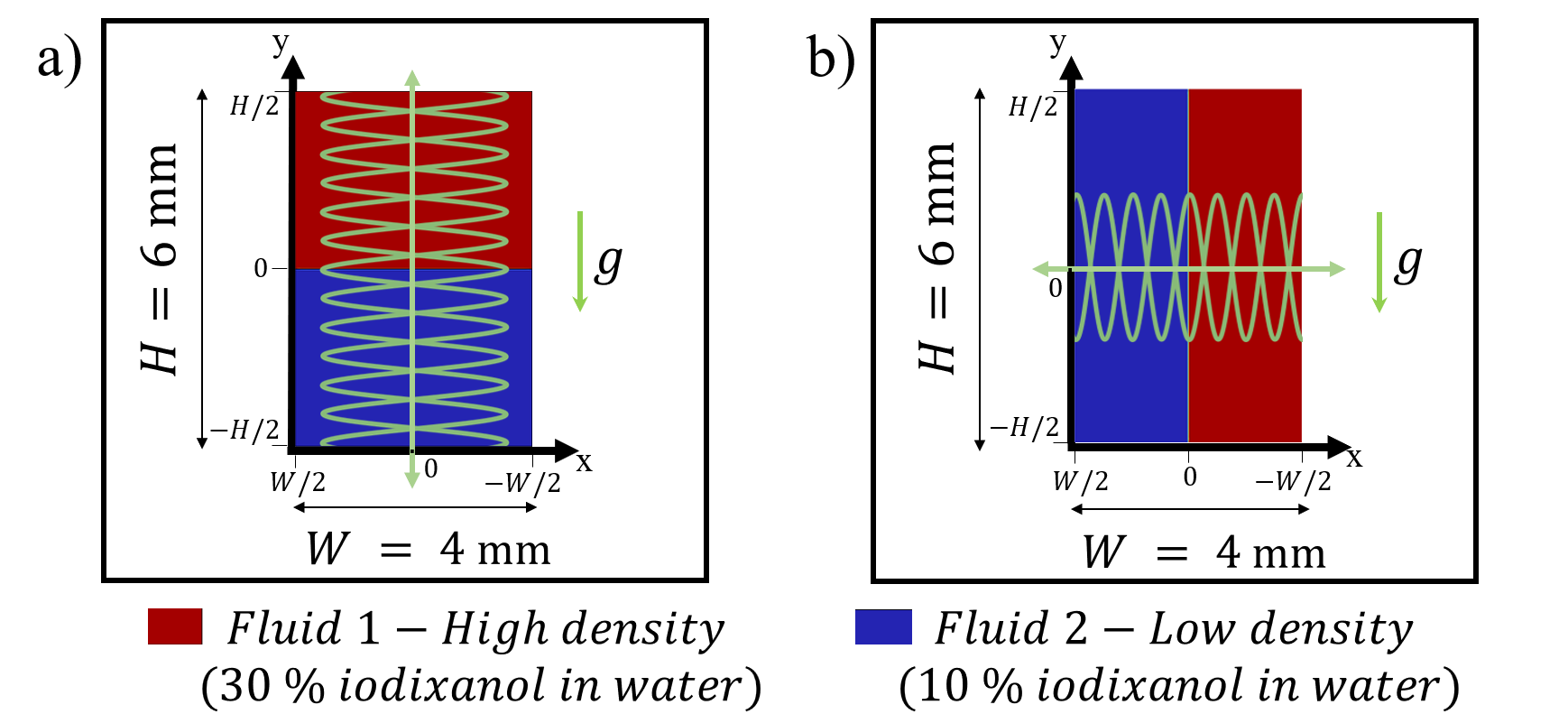}
    \caption{Schematic representation of the initial Configurations. (a) Configuration I: high-density fluid in the top half and low-density fluid in the bottom half. (b) Configuration II: high-density fluid in the right half and low-density fluid in the left half of the channel. Standing bulk acoustic waves (represented by green wave patterns) are depicted propagating perpendicular to the fluid interface, while gravity ($g$) acts in the negative $y$-direction.}
\label{sch}
\end{figure}

This theoretical study examines the interplay between gravitational and acoustic forces on stratified miscible fluids. To demonstrate how acoustic fields can suppress gravity-induced fluid mixing, two different configurations are investigated: horizontal (Configuration I) and vertical (Configuration II) fluid arrangements, as depicted in Figs.~\ref{sch}(a) and (b), respectively. Configuration I starts in an unstable gravitational equilibrium susceptible to the Rayleigh-Taylor instability (RTI), whereas Configuration II begins in a state of gravitational non-equilibrium.

\subsection{{\textbf{Governing Equations}}}

The spatiotemporal evolution of inhomogeneous miscible acoustofluidic systems is governed by the fundamental principles of mass conservation, momentum conservation, and advection-diffusion. These governing equations are expressed respectively as:\cite{Landau1987Aug}
\begin{subequations}
\begin{equation}
\label{eq1a}
\frac{\partial \rho}{\partial t}+\nabla \cdot(\rho \boldsymbol{V})=0,
\end{equation}
\begin{equation}
\label{eq1b}
\rho \frac{\mathrm{D} \boldsymbol{V}}{\mathrm{D} t}=-\nabla p+\eta \nabla^2 \boldsymbol{V}+\beta \eta \nabla(\boldsymbol{\nabla} \cdot \boldsymbol{V})+\boldsymbol{f}_{ac_2} + \rho\boldsymbol{g},
\end{equation}
\begin{equation}
\label{eq1c}
\partial_t \phi+\boldsymbol{V}\cdot\nabla \phi = D\nabla^2 \phi,
\end{equation}
\end{subequations}
where $\rho$ is the fluid density, $\boldsymbol{V}$ the velocity field, $p$ the pressure, $\eta$ the dynamic viscosity, $\beta = (\xi / \eta) + (1 / 3)$ with $\xi$ representing bulk viscosity, and $\boldsymbol{f}_{ac}$ is the acoustic body force term. $\mathrm{D} / \mathrm{D} t=\partial_t+\boldsymbol{V} \cdot \nabla$ denotes the material derivative, $\boldsymbol{g}$ is the acceleration due to gravity in the negative $y$-direction, $\phi$ is the solute concentration, and $D$ is the diffusivity. In this study, both fluids are considered incompressible; hence, Eq.~\ref{eq1a} reduces to $\boldsymbol{\nabla} \cdot \boldsymbol{V}=0$, and the term $\beta \eta \nabla(\boldsymbol{\nabla} \cdot \boldsymbol{V})$ in Eq.~\ref{eq1b} vanishes.

The physics of the problem involves two distinct timescales. The primary acoustic fields ($\omega \sim 2\ \mathrm{MHz}$) vary on a fast timescale ($\sim 0.1\ \mathrm{\mu s}$), whereas the induced hydrodynamic responses, such as acoustic relocation, streaming, and radiation forces, evolve on a slower timescale ($\sim 1\ \mathrm{ms}$). Our numerical model explicitly solves only the slow-scale fields (Eqs.~\ref{eq1a}, \ref{eq1b}, and \ref{eq1c}), while the first-order fast-scale pressure ($\boldsymbol{p_1}$) and velocity ($\boldsymbol{v}_1$) of the acoustic standing wave are analytically assumed. The fast-time acoustic fields drive the slow-time scale phenomena through the acoustic body force ($\boldsymbol{f}_{ac}$), which originates from the divergence of the time-averaged Reynolds stress tensor of the primary acoustic fields and is mathematically expressed as:\cite{KarlsenAugustssonBruus2016}
\begin{equation}
\label{eq2}
\boldsymbol{f}_{ac} = -\nabla \cdot \langle \rho \boldsymbol{v}_1 \otimes \boldsymbol{v}_1 \rangle,
\end{equation}
where $\otimes$ denotes the tensor product. By neglecting acoustic streaming effects, the Eq.~\ref{eq2} simplifies to:\cite{Rajendran2022}
\begin{equation}
\label{eq4}
\begin{aligned}
\boldsymbol{f}_{ac} &= \frac{1}{2}\nabla\left(\kappa \langle|p_1|^2\rangle - \rho \langle|\boldsymbol{v}_1|^2\rangle\right) \\
&- \frac{1}{2}\left(\langle|p_1|^2\rangle \nabla \kappa + \langle|\boldsymbol{v}_1|^2\rangle \nabla \rho\right) = \nabla \phi_{1} + \boldsymbol{f}_{rl},
\end{aligned}
\end{equation}
where $\kappa = 1 / (\rho c^2)$ is the fluid compressibility, and the angle brackets $\langle \dots \rangle$ indicate time averaging over one oscillation period of the wave. For a standing wave propagating in the $y$-direction, the fast-scale acoustic pressure ($p_1$) and velocity ($\boldsymbol{v}_1$) are defined as follows:\cite{KarlsenAugustssonBruus2016}
\begin{subequations}
\begin{equation}
\label{eq3a}
\boldsymbol{p}_1= p_a \sin(k y), 
\end{equation}
\begin{equation}
\label{eq3b}
\boldsymbol{v}_1= \frac{p_a}{i \rho c} \cos(ky), 
\end{equation} 
\end{subequations}
where $p_a$ is the pressure amplitude, $c$ is the speed of sound in the medium, $k = 2\pi/\lambda$ is the wavenumber, and $\lambda$ is the acoustic wavelength. 

In Eq.~\ref{eq4}, the gradient term ($\nabla \phi_{1}$) is absorbed into the pressure field and does not contribute to the velocity. Consequently, the acoustic relocation is driven entirely by the non-gradient component of the body force ($\boldsymbol{f}_{rl}$), which evaluates to:\cite{KarlsenAugustssonBruus2016}
\begin{equation}
\label{eq5}
\boldsymbol{f}_{rl} = - \left[\frac{1}{4} |p_1|^2 \nabla \kappa + \frac{1}{4}|\boldsymbol{v}_1|^2 \nabla \rho\right].
\end{equation}

For a standing acoustic wave in the $y$-direction (as shown in Fig.~\ref{sch}(a)) with acoustic pressure (Eq.~\ref{eq3a}) and velocity (Eq.~\ref{eq3b}), the above equation (Eq.~\ref{eq5}) can be approximated and simplified to: \cite{Rajendran2022,Thirisangu2025UnifiedDroplets}
\begin{subequations}
\begin{align}
\label{eq6a}
    \boldsymbol{f}_{ac_1} &= -E_{ac} \cos(2ky) \nabla \hat{Z}, \\
    &= -\nabla (E_{ac} \cos(2ky) \hat{Z}) - 2kE_{ac} \sin(2ky) \hat{Z} \mathbf{e}_y, \nonumber \\
    &= -\nabla\phi_{2} - 2kE_{ac} \sin(2ky) \hat{Z} \mathbf{e}_y, \nonumber
\end{align} 
where $E_{ac} = p_a^2 / (4 \rho_{avg} c_{avg}^2)$ is the acoustic energy density and $Z = \rho c$ denotes the acoustic impedance. The normalized parameters are given by $\hat{Z} = Z / Z_{avg}$, $\hat{c} = c / c_{avg}$, and $\hat{\rho} = \rho / \rho_{avg}$, where the subscript 'avg' denotes the respective average quantities of the two selected fluids. $\mathbf{e}_y$ is
the unit normal vector along the $y$-direction and $\mathbf{e}_x$ is
the unit normal vector along the $x$-direction. In closed system, the gradient term ($-\nabla\phi_{2}$) is absorbed into the pressure field and it does not induce bulk fluid flow. Consequently, the non-gradient term is solely responsible for driving the acoustic relocation.\cite{Thirisangu2025UnifiedDroplets} Therefore, the acoustic body force can be expressed as:
\begin{equation}
\label{eq6b}
    f_{ac_2} = -2kE_{ac} \sin(2ky)\hat{Z}\mathbf{e}_y.
\end{equation}
\end{subequations}

Equations \ref{eq6a} and \ref{eq6b} can be used interchangeably.\cite{Thirisangu2025UnifiedDroplets} The only difference between them lies in the renormalization of the pressure field, which has no influence on fluid flow or mixing within a closed system; the same applies to the following Equations \ref{eq7a} and \ref{eq7b} as well. 
For a standing acoustic wave in the $x$-direction (as shown in Fig.~\ref{sch}(b)), with fast-scale acoustic pressure ($\boldsymbol{p}_1= p_a \sin(kx)$) and velocity ($\boldsymbol{v}_1= \frac{p_a}{i \rho c} \cos(kx)$), the acoustic body force takes the form:
\begin{subequations}
\begin{equation}
    \label{eq7a}
\boldsymbol{f}_{ac_1} = -E_{ac} \cos(2k x) \nabla \hat{Z}.
\end{equation}
Similarly, the above equation can be expressed as:
\begin{equation}
\label{eq7b}
f_{ac_2} = -2kE_{ac} \sin(2kx)\hat{Z}\mathbf{e}_x.
\end{equation}
\end{subequations}
Importantly, in the acoustic body force equations (Eqs.~\ref{eq6a} and \ref{eq7a}), the impedance gradient ($\nabla \hat{Z}$) plays a pivotal role in the acoustic relocation mechanism. In the theoretical framework established in our previous microchannel study,\cite{Rajendran2022} gravitational effects were neglected. Extending this framework, the present study focuses on slow-scale fluid phenomena in a minichannel, which are fundamentally governed by the dynamic interplay between the acoustic body force and gravity.

\subsection{{\textbf{Quantitative Evaluation of Mixing Suppression}}}

To numerically quantify the efficiency of the proposed acoustofluidic suppression method, we evaluate the mixing index (MI),\cite{Pothuri2019Dec} formulated mathematically as:

\begin{equation}
\begin{aligned}
M I=1-\frac{\sigma_i}{\sigma_0}, \text { where } \sigma_i & =\sqrt{\frac{1}{N} \sum_{i=1}^N\left(\phi_i-\phi_m\right)^2}, \\
\sigma_o & =\sqrt{\frac{1}{N} \sum_{i=1}^N\left(\phi_o-\phi_m\right)^2},
\end{aligned}
\end{equation}
The point concentration at  $i^{th}$ sample point is denoted by $\phi_i$ and $\phi_m$ is the mean concentration, $\phi_o$ is the initial concentration of the non-mixed section, $\sigma$ is the standard deviation of the concentration in a given cross section, and $N$ is the total number of sampling points. The mixing index (MI) value lies between 0 and 1. A mixing index of value 1 (MI = 1) indicates that the fluids are completely mixed (homogeneous mixture) and a mixing index of value 0 (MI = 0) indicates that the fluids are completely unmixed (no mixing occurred). This quantifiable parameter allows us to accurately assess the inhomogeneous state of the fluid mixture and evaluate the efficiency of mixing suppression.

\subsection{{\textbf{Numerical method}}}

The computational domain comprises a rectangular minichannel with a cross-sectional width ($W$) of 4 mm in the $x$-direction and a height ($H$) of 6 mm in the $y$-direction (Fig.\ref{sch}). Aqueous iodixanol solutions at concentrations of 10\% and 30\% are selected as the working fluids to establish the necessary acoustic impedance contrast. The physical properties of these fluids, including density, viscosity, and speed of sound in the fluid medium, are parameterized in terms of the solute concentration as given in the literature (see Table~\ref{tab:1} in Appendix~\ref{appendix:A}). The governing continuity (Eq.~\ref{eq1a}), momentum (Eq.~\ref{eq1b}), and advection-diffusion (Eq.~\ref{eq1c}) equations along with acoustic body force equation (Eq.~\ref{eq6b} for wave application in $y$-direction or Eq.~\ref{eq7b} for wave application in $x$-direction) are solved using COMSOL Multiphysics (versions 6.2 and 6.3). For fixed fluid properties and wavelength ($\lambda$), the acoustic body force is varied by varying the acoustic energy density ($E_{ac}$). The Laminar Flow module is utilized to define the fluid properties and boundary conditions, and to integrate the acoustic radiation force as a body force term. Concurrently, the Transport of Diluted Species module is utilized to specify the diffusion coefficient (see Appendix~\ref{appendix:A}), initialize the spatial positioning of the fluids, and define their respective concentrations. This time-dependent study is solved using the Implicit Backward Differentiation Formula, where the dependent variables (velocity field, pressure, and concentration) are computed using a segregated Parallel Direct Solver (PARDISO). The fluid domain is initialized in a stratified, perfectly quiescent state ($\mathbf{V} = 0$ at $t = 0$), with standard no-slip boundary conditions enforced along all channel walls. 

\begin{figure}[h]
\centering
\includegraphics[scale=0.56]{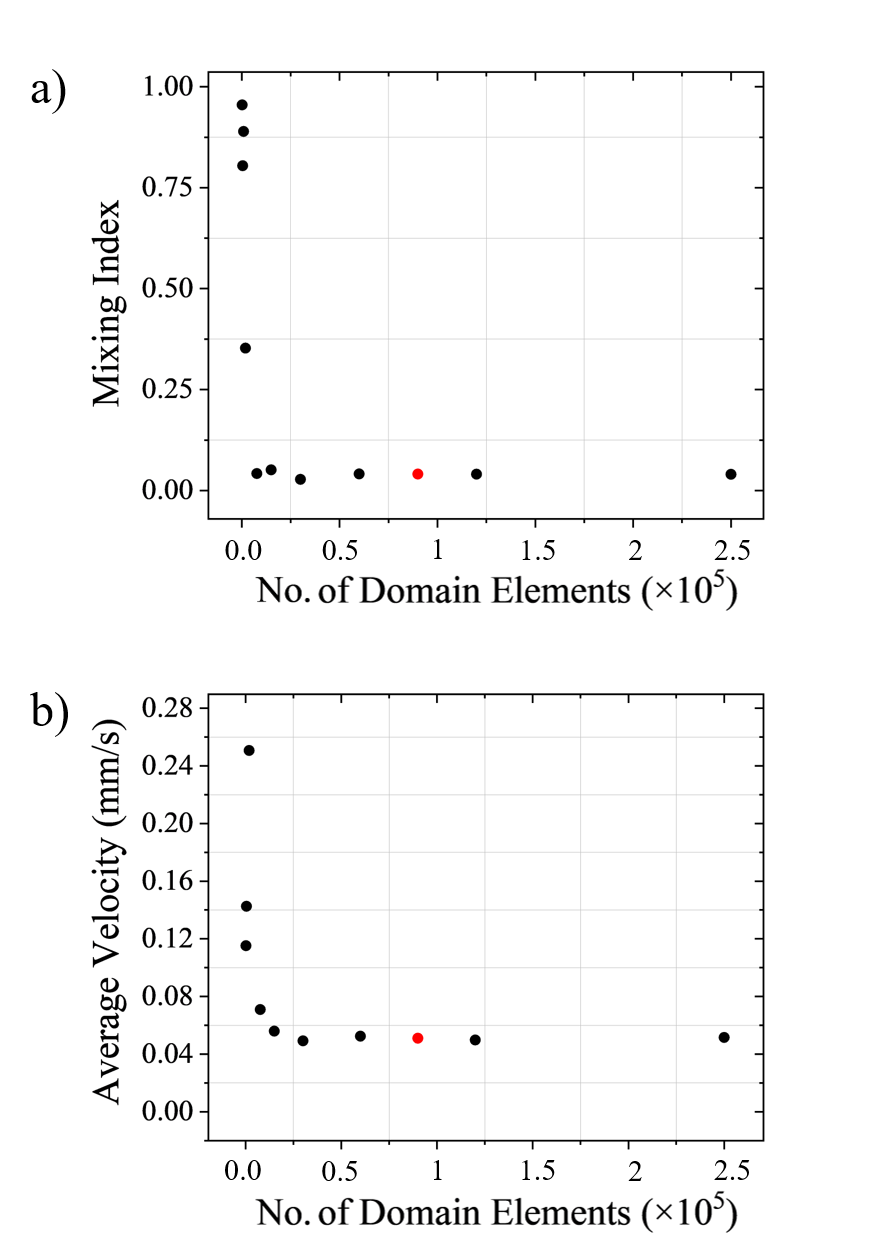}
\caption{Grid independence study illustrating the variation of (a) Mixing Index and (b) average velocity with respect to the number of domain elements. The analysis was conducted in Configuration I, where the standing acoustic waves (with $E_{ac} = 125$ J/m$^3$ and $\lambda = 0.3$ mm) are oriented perpendicular to the fluid–fluid interface. The red marker denotes the selected mesh resolution (90,722 elements), ensuring a sufficiently fine computational grid to achieve a grid-independence solution.}
\label{GID}
\end{figure}

In idealized numerical simulations, a perfectly flat interface creates an unstable equilibrium, artificially delaying the onset of the Rayleigh-Taylor Instability (RTI) until computational errors arbitrarily break the symmetry. To prevent this computational stagnation and deterministically capture the RTI evolution in Configuration I, an initial downward perturbation of 0.001 mm is strategically seeded at the interface midpoint. 
Importantly, the acoustic field is configured such that the fluid-fluid interface lies exactly midway between a pressure node and an antinode (see Fig.~\ref{rel1}(a)(iii)), thereby subjecting it to the maximum acoustic radiation force. For achieving this in Configuration I, where the interface is located at $y = 0$ (Fig.~\ref{sch}(a)), a phase shift of $\pi/2$ is introduced into Eq.~\ref{eq6b}, yielding an acoustic body force expression: $f_{ac_2} = -2kE_{ac} \sin(2ky+\pi/2)\hat{Z}\mathbf{e}_y$. To ensure grid independence, a mesh convergence analysis (Fig.~\ref{GID}) was performed, leading to the selection of a computational grid comprising 90,722 elements (maximum element size of 0.026 mm). As denoted by the red marker in Fig.~\ref{GID}, this optimal mesh configuration achieves numerical convergence and is utilized for all subsequent simulations in this study.

In our previous investigations, the present computational framework has been validated across diverse acoustofluidic phenomena, such as acoustic relocation of inhomogeneous fluids,\cite{Rajendran2022} streaming suppression,\cite{Rajendran2022} and droplet manipulation (migration, splitting and deformation).\cite{Thirisangu2025UnifiedDroplets} In these studies, the numerical methodology was benchmarked against established experimental studies focusing on acoustic radiation forces at liquid interfaces \cite{DeshmukhBrzozkaLaurellAugustsson2014} and the relocation of coflowing immiscible liquids.\cite{HemachandranKarthickLaurellSen2019} Deploying this previously validated numerical framework ensures the predictive accuracy of our current simulations.

\section{\textbf{Results and discussion}}\label{sec3}

This section demonstrates an acoustofluidic strategy to inhibit the Rayleigh-Taylor instability (RTI) and suppress gravity-driven fluid mixing. By examining the mixing dynamics governed by the interplay of gravitational and acoustic forces, we systematically analyze how variations in acoustic energy density, wavelength, and wave orientation dictate the stability of stratified fluids subjected to an acoustic field in a minichannel.

\subsection{\textbf{Acoustofluidic suppression of mixing in Configuration I: High-density fluid above a low-density fluid}}

\begin{figure*}[p!]
    \centering
     \includegraphics[
        width=\textwidth,
        height=\textheight,
        keepaspectratio
    ]{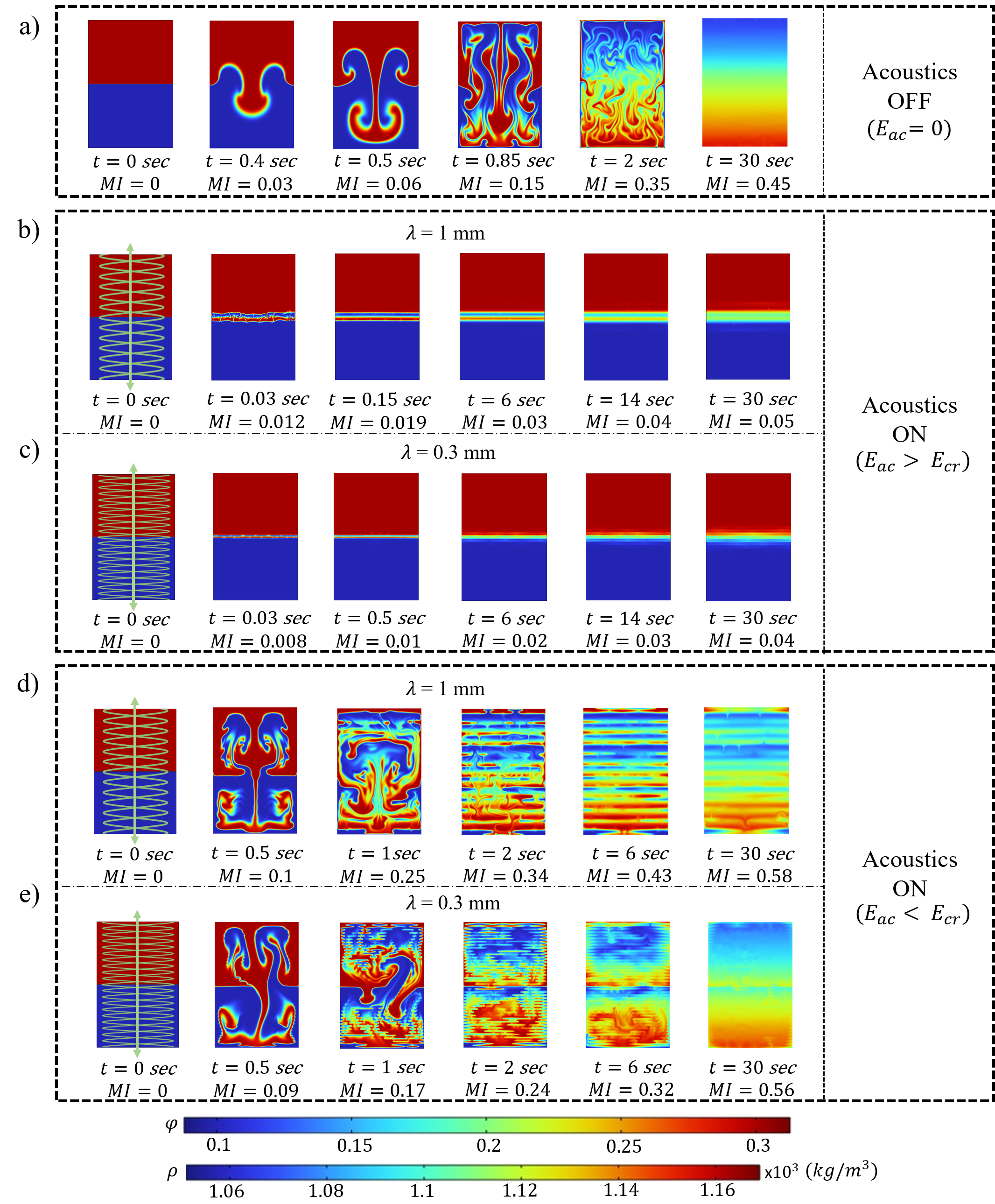}
    
    \caption{Time evolution of concentration fields illustrating the mixing dynamics for Configuration I. The critical acoustic energy densities are $E_{cr} = 5$ J/m$^3$ for $\lambda = 1$ mm and $E_{cr} = 4.6$ J/m$^3$ for $\lambda = 0.3$ mm. (a) Baseline mixing condition ($E_{ac} = 0$), where mixing is induced solely by gravity. (b, c) Acoustically dominated regime ($E_{ac} > E_{cr}$), with standing bulk acoustic waves applied perpendicular to the fluid interface for $\lambda = 1$ mm and $\lambda = 0.3$ mm, respectively, both maintained at an acoustic energy density of 125 J/m$^3$. (d, e) Gravity-dominated regime ($E_{ac} < E_{cr}$), with standing bulk acoustic waves applied perpendicular to the fluid interface for $\lambda = 1$ mm and $\lambda = 0.3$ mm, maintained at applied acoustic energy densities of 1.75 J/m$^3$ and 2.5 J/m$^3$, respectively.}
    \label{Case1}
\end{figure*}

Figures \ref{Case1}(a)-(e) present the mixing dynamics of Configuration I (Fig.~\ref{sch}(a), gravitationally unstable equilibrium) under different conditions. The baseline condition ($E_{ac} = 0$) is illustrated in Fig.~\ref{Case1}(a), where, in the absence of an acoustic field, the fluid behavior is governed entirely by gravitational forces and molecular diffusion. The Rayleigh-Taylor instability (RTI) initiates as a cup-shaped interfacial deformation at $t = 0.4$ s and rapidly evolves into a classic mushroom-shaped morphology by $t = 0.5$ s. As the instability amplifies, chaotic fluid mixing takes place (as shown in Fig.~\ref{Case1}(a) at $t = 2$ s). Ultimately, the fluids settle into the gravitationally stable equilibrium, with the high-density fluid at the bottom and the low-density fluid at the top, leaving subsequent homogenization to be governed solely by molecular diffusion (Fig.~\ref{Case1}(a)) and yielding a mixing index of 0.45 at $t = 30$ s.

As illustrated in Fig.~\ref{Case1}(b), the acoustic body force (with $E_{ac}>E_{cr}$) dominates gravity, completely inhibiting the Rayleigh-Taylor instability and suppressing the gravity-driven fluid mixing. By operating above the critical threshold ($E_{ac} > E_{cr}$), the acoustic body force immediately pins the interface via acoustic relocation (as shown in Fig.~\ref{Case1}(b) at $t = 0.03$ s). This establishes an acoustically controlled barrier (as shown in Fig.~\ref{Case1}(b) at $t = 0.15$ s), which prevents gravitationally induced convective flow. With gravity-induced fluid convection suppressed, fluid homogenization occurs only due to slow molecular diffusion between the relocated fluid layers (Fig.~\ref{Case1}(b)). Consequently, this yields a significantly reduced mixing index (MI) of 0.05 at $t = 30$ s for an acoustic wavelength of $\lambda = 1$ mm, compared to an MI of 0.45 (at $t = 30$ s) obtained in the baseline condition ($E_{ac} = 0$, mixing induced solely by gravity). Similarly, Fig.~\ref{Case1}(c) demonstrates identical stabilizing phenomena using a shorter acoustic wavelength of 0.3 mm. As this reduced wavelength decreases the spatial distance between the pressure nodes and antinodes, it restricts interfacial deformation even further, resulting in a relatively lower MI of 0.04 at $t = 30$ s.

\begin{figure*}[!t] 
    \centering
    \includegraphics[width=0.82\linewidth]{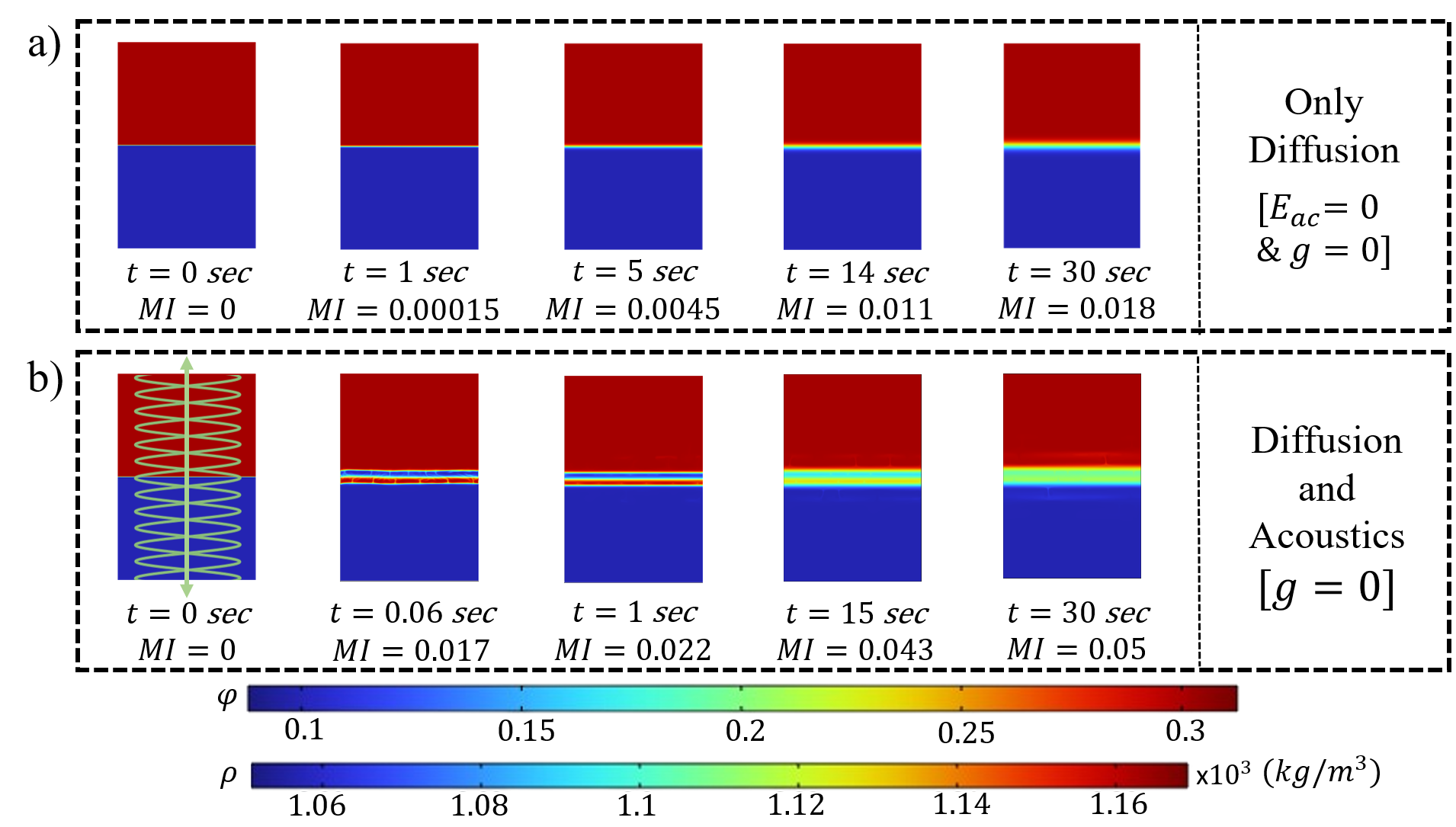}
    \caption{Time evolution of the concentration fields illustrating the mixing dynamics for Configuration I in the absence of gravity. (a) Mixing only due to diffusion (idealized baseline mixing condition), where both acoustics and gravity are absent ($E_{ac} = 0$ and $g = 0$). (b) Mixing only due to acoustics and diffusion ($g = 0$), where standing bulk acoustic waves with an acoustic energy density $E_{ac} = 125$ J/m$^3$ are applied perpendicular to the fluid interface at a wavelength of $\lambda = 1$ mm.}
    \label{only_diff_and_acoustics}
\end{figure*}
In contrast to the stabilizing condition discussed above ($E_{ac} > E_{cr}$, Figs.~\ref{Case1}(b) and (c)), reducing the applied acoustic energy density below the critical threshold ($E_{ac} < E_{cr}$) triggers a counterintuitive system response: fluid mixing is enhanced, with the mixing index (MI) increasing to a level that exceeds the gravity-driven baseline. Figures~\ref{Case1}(d) and (e) illustrate the fluid behavior for this sub-critical case ($E_{ac} < E_{cr}$), demonstrating this adverse effect. As the applied acoustic field is insufficient to sustain the relocated fluids at the interface, acoustic interfacial pinning against gravity is no longer possible. Consequently, the dominant gravitational force induces fluid convection (as shown in Fig.~\ref{Case1}(d) at $t = 0.5$ s). However, the weak acoustic force still attempts to relocate these dispersing fluids, fragmenting them into multiple distinct bands throughout the fluid domain (as shown in Fig.~\ref{Case1}(d) at $t = 2$ s). As these bands increase the number of fluid-fluid interfaces, this acoustic effect acts alongside gravitational convection to drive an enhancement in fluid mixing, yielding an elevated mixing index (MI) of 0.58 at $t = 30$ s for an acoustic wavelength of $\lambda = 1$ mm.  A similar system response is observed for a shorter acoustic wavelength of $\lambda = 0.3$ mm, also yielding an elevated MI of 0.56 at $t = 30$ s. The underlying physical mechanisms and the interplay of forces governing these behaviors will be detailed in the following sections.

While the results under the stabilization condition ($E_{ac} > E_{cr}$, Figs.~\ref{Case1}(b) and (c)) highlight the capacity of the acoustic body force to arrest gravity-driven fluid convection and suppress the resulting mixing, it is crucial to recognize that the acoustic field cannot prevent mixing driven by molecular diffusion. Figure~\ref{only_diff_and_acoustics}(a) presents the idealized baseline mixing condition where both gravitational and acoustic fields are absent ($E_{ac} = 0$ and $g = 0$), leaving fluid mixing to be governed solely by slow molecular diffusion. Under this idealized state, the system yields a mixing index (MI) of just 0.018 at $t = 30$ s. Further, to isolate the effect of the acoustic field, a zero-gravity condition subjected solely to acoustic forces is illustrated in Fig.~\ref{only_diff_and_acoustics}(b). Here, acoustic relocation between local nodes and antinodes disrupts interfacial homogeneity, yielding an MI of 0.05 at $t = 30$ s, which exceeds the idealized baseline MI of 0.018. These results highlight a fundamental physical limit of this suppression mechanism: while the acoustic field successfully arrests gravity-driven convection and completely inhibits the Rayleigh-Taylor instability, fluid mixing cannot be suppressed below the idealized baseline, where mixing occurs solely due to diffusion. As the molecular diffusion between the acoustically relocated fluid layers is inevitable, regardless of the effectiveness of interfacial pinning, the idealized scenario (Fig.~\ref{only_diff_and_acoustics}(a)) establishes the absolute lower bound for mixing. However, as demonstrated in the preceding results (Fig.~\ref{Case1}(b) and (c)), the acoustic field substantially suppresses convective fluid mixing induced by the gravitational force, reducing the mixing index from MI = 0.45 to MI = 0.05 at $t = 30$ s (Fig.~\ref{Case1}(a) and (b)).

\subsubsection{Acoustically dominated regime ($E_{ac} > E_{cr}$)}

\begin{figure*}[t!]
\centering
\includegraphics[width=1\linewidth]{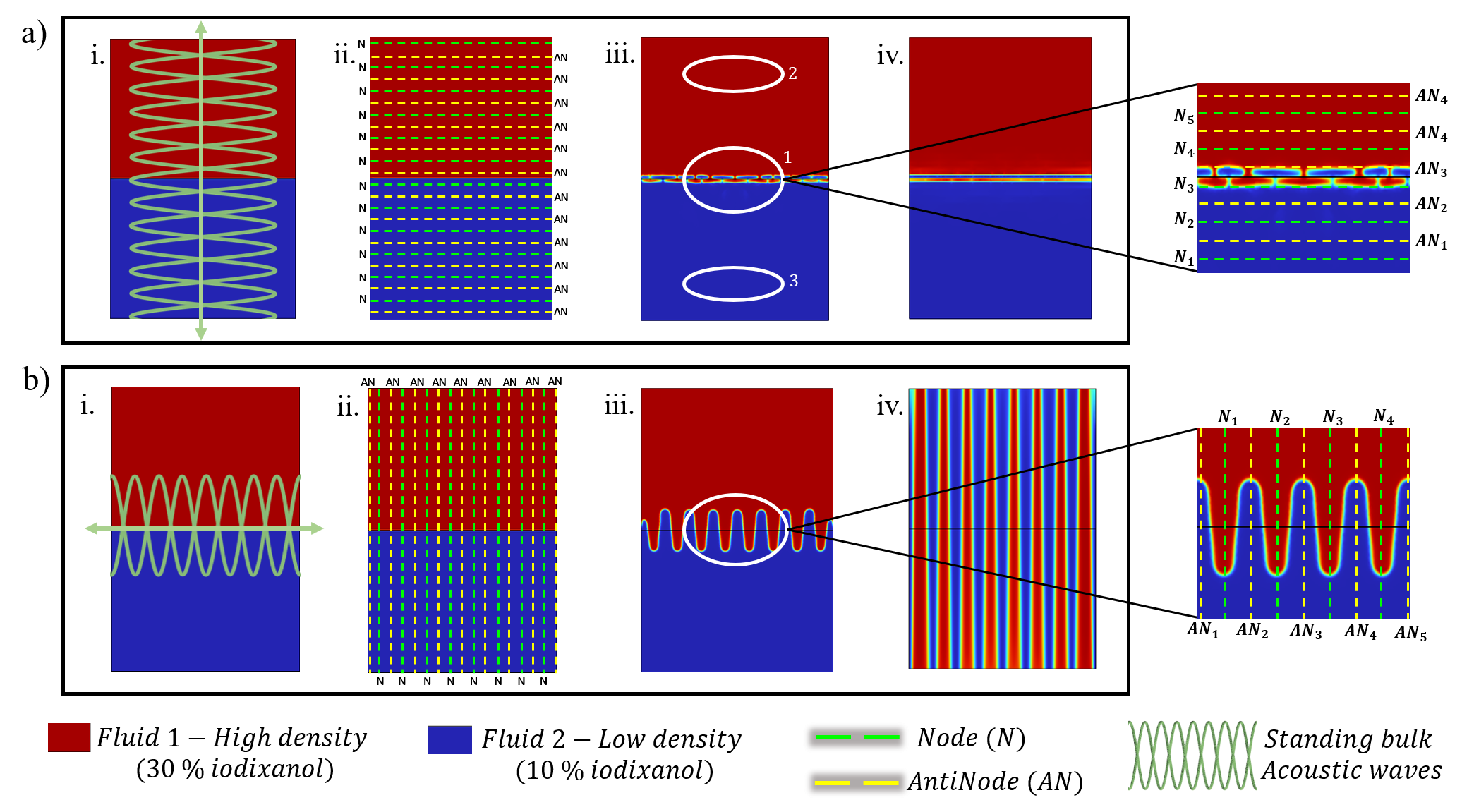}
\caption{Schematic representation of the acoustic relocation for Configuration I under different standing wave orientations in the acoustically dominated regime ($E_{ac} > E_{cr}$). (a) Fluid relocation behavior when standing acoustic waves are applied perpendicular to the fluid-fluid interface. (b) Fluid relocation behavior when standing acoustic waves are applied parallel to the fluid-fluid interface. (The green dotted lines indicate the nodes, whereas the yellow dotted lines indicate the antinodes).} 
\label{rel1}
\end{figure*}

This section details the physical mechanisms driven by the acoustic body force in the acoustically dominated regime ($E_{ac} > E_{cr}$) and examines the effect of wave propagation direction on fluid mixing. During acoustic resonance, the interaction between incident and reflected waves generates standing bulk acoustic waves (BAWs). These acoustic waves create pressure nodes and antinodes across the fluid domain, as illustrated in Figs.~\ref{rel1}(a)(ii) and \ref{rel1}(b)(ii).

The resulting acoustic radiation force drives a spatial reorganization of the fluids, forcing the high-impedance fluid to migrate toward the nearest pressure node and the low-impedance fluid toward the nearest antinode (as shown in Figs.~\ref{rel1}(a)(iii) and \ref{rel1}(b)(iii)). With the acoustic relocation phenomenon outlined, it is necessary to examine how the standing wave direction governs the stability of the fluid system. An evaluation of different wave orientations demonstrates that in configurations where gravitational instabilities are present, stabilization can be achieved only when the acoustic waves are applied perpendicular to the fluid interface. In contrast, when the standing waves are applied parallel to the fluid interface, mixing is significantly enhanced.\cite{Pothuri2019Dec} To understand these contrasting system behaviors, the localized physical mechanisms driving each of the flow behaviors are detailed below.

First, considering the acoustic stabilization/suppression condition: for an initially horizontal interface, waves propagating perpendicular to the fluid-fluid interface (the $y$-direction) create acoustic nodes and antinodes across the fluid domain, as shown in Fig.~\ref{rel1}(a)(ii). Driven by the acoustic radiation force ($f_{ac_2} = -2kE_{ac} \sin(2ky)\hat{Z}\mathbf{e}_y$), the fluids at the interface relocate to its nearest stable positions (the high-impedance fluid migrates to the nearest node, and the low-impedance fluid to the nearest antinode, as shown in Fig.~\ref{rel1}(a)(iii)). Under this condition, the interface is spatially bounded by only a single adjacent node and antinode, which confines the fluid relocation to only these two specific acoustic regions. In this acoustically dominated regime ($E_{ac} > E_{cr}$), once relocated, the fluids are trapped in their respective acoustically stable regions (high-density fluid at the nodes and low-density fluid at the antinodes). This prevents the fluids traversing from one node to another (which requires bypassing an antinode) or from one antinode to another (which requires bypassing a node). Consequently, the interface becomes acoustically pinned against gravity (as shown in Fig.~\ref{rel1}(a)(iv)). Ultimately, this pinned interface acts as a barrier that successfully inhibits the Rayleigh-Taylor instability and effectively suppresses gravity-driven bulk fluid mixing.  As the acoustic radiation force requires an impedance ($Z$) gradient ($\nabla \hat{Z} \neq 0$) to induce motion, it has no effect on the bulk homogeneous fluid regions (Fig.~\ref{rel1}(a)(iii), marked as regions 2 and 3); fluid relocation is exclusively confined to the inhomogeneous region (Fig.~\ref{rel1}(a)(iii), marked as region 1). Importantly, while the acoustic force completely prevents gravity-driven bulk movement, slow molecular diffusion still prevails. Over extended periods, the fluid mixing due to diffusion, which occurs layer by layer, expands the localized impedance gradient ($\nabla \hat{Z} \neq 0$) into the initially homogeneous bulk domains, eventually leading to complete fluid homogenization. Nevertheless, this acoustofluidic approach provides a substantial temporal window during which the gravitationally induced fluid mixing is effectively suppressed. Crucially, if gravitational convection is allowed to initiate prior to acoustic actuation, the resulting fluid dispersion becomes irreversible. 
\begin{figure*}[p!]
    \centering
 \includegraphics[
        width=\textwidth,
        height=\textheight,
        keepaspectratio
    ]{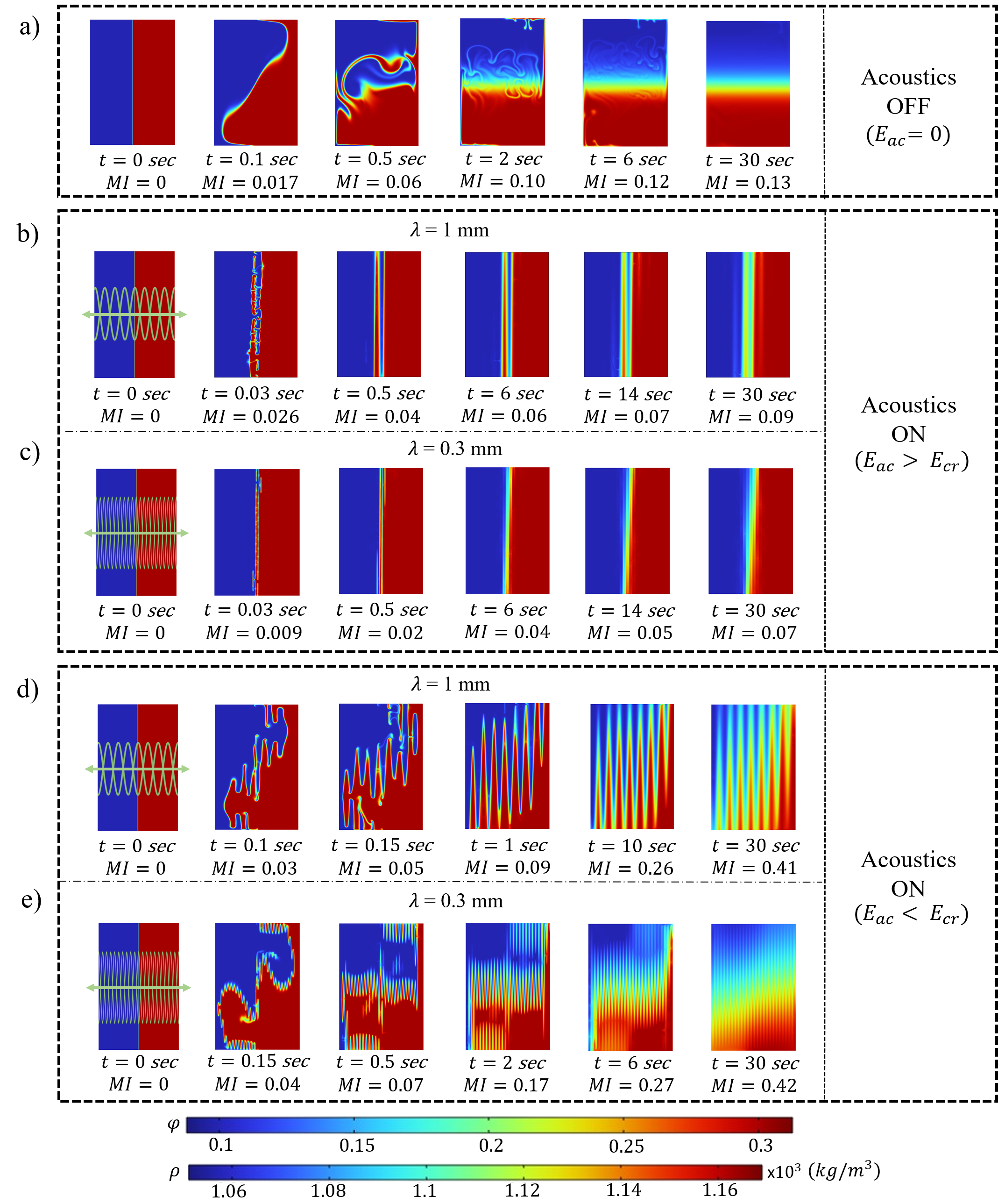}
    \caption{Time evolution of concentration fields illustrating the mixing dynamics for Configuration II. The critical acoustic energy densities are $E_{cr} = 75$ J/m$^3$ for $\lambda = 1$ mm and $E_{cr} = 62.5$ J/m$^3$ for $\lambda = 0.3$ mm. (a) Baseline mixing condition ($E_{ac} = 0$), where mixing is induced solely by gravity. (b, c) Acoustically dominated regime ($E_{ac} > E_{cr}$), with standing bulk acoustic waves applied perpendicular to the fluid interface for $\lambda = 1$ mm and $\lambda = 0.3$ mm, respectively, both maintained at an acoustic energy density of 125 J/m$^3$. (d, e) Gravity-dominated regime ($E_{ac} < E_{cr}$), with standing bulk acoustic waves applied perpendicular to the fluid interface for $\lambda = 1$ mm and $\lambda = 0.3$ mm, maintained at applied acoustic energy densities of 35 J/m$^3$ and 32.5 J/m$^3$, respectively.}   
    \label{Case2}
\end{figure*}
Once the fluids are actively displaced, even an acoustic energy density exceeding the critical threshold ($E_{ac} > E_{cr}$) is incapable of restabilizing the interface or reversing the gravitationally induced fluid mixing. Thus, successful interfacial pinning requires acoustic actuation from the initial state ($t = 0$).

Second, considering the alternative wave orientation: for an initially horizontal interface, when acoustic standing waves propagate parallel to the fluid interface (the $x$-direction), the fluid response differs significantly from that of the perpendicular wave application condition (as seen in Fig.~\ref{rel1}(a)). Here, the fluid interface is in direct contact with all the pressure nodes and antinodes generated by the standing bulk acoustic waves (as shown in Fig.~\ref{rel1}(b)(ii)). Consequently, the fluid migration toward its nearest available stable acoustic regions results in the distinct finger-like patterns (as shown in Fig.~\ref{rel1}(b)(iii)). While operating in the acoustically dominated regime ($E_{ac} > E_{cr}$), this relocation forces the fluids to stratify into alternating vertical columns against gravity (as shown in Fig.~\ref{rel1}(b)(iv)), which increase the number of fluid-fluid interfaces. As these multiple stratified layers amplify the available area for molecular diffusion, an enhancement in fluid mixing is observed. Therefore, applying acoustic waves parallel to the fluid interface must be strictly avoided when the primary objective is to suppress the instability and the resulting gravity-induced fluid mixing.

\subsection{\textbf{Acoustofluidic suppression of fluid mixing in Configuration II: High‑ and low‑density fluids placed at the right and left halves of the channel, respectively}}

Following the analysis of Configuration I, the mixing dynamics of Configuration II are now examined across three different conditions, as shown in Fig.~\ref{Case2}. To establish the baseline mixing condition, the gravity-induced mixing without the influence of acoustics ($E_{ac} = 0$) is evaluated first (Fig.~\ref{Case2}(a)). Here, the initial vertical stratification of the fluids under gravity creates a non-equilibrium condition. As a result, the system's high potential energy is rapidly converted into kinetic energy, driving the high-density fluid diagonally downward (as shown in Fig.~\ref{Case2}(a) at $t = 0.1$ s). Subsequently, the inertia of the fluid causes a transient hydrodynamic overshoot, resulting in an upward rebound (as shown in Fig.~\ref{Case2}(a) at $t = 0.5$ s). Further, viscous dissipation decays momentum and enables the system to settle into a stable, horizontally stratified equilibrium (as shown in Fig.~\ref{Case2}(a) at $t = 2$ s). Thereafter, mixing of fluids proceeds exclusively via slow molecular diffusion (Fig.~\ref{Case2}(a) at $t = 6$ s). Due to the absence of the chaotic mixing inherent to Configuration I, the baseline condition ($E_{ac} = 0$) for Configuration II yields a significantly lower mixing index (MI = 0.13) compared to Configuration I (MI = 0.45) at $t = 30$ s (Figs.~\ref{Case1}(a) and \ref{Case2}(a)).

Building upon this baseline, Figs.~\ref{Case2}(b) and (e) illustrate the mixing dynamics under acoustic forcing. Despite the inherent non-equilibrium state of configuration II, the acoustically dominated regime ($E_{ac} > E_{cr}$) successfully arrests gravity-induced bulk convection via the same stabilization mechanisms established for Configuration I. Consequently, fluid mixing proceeds only through molecular diffusion across the acoustically pinned interface, yielding a mixing index (MI) of 0.09 with $\lambda = 1$ mm (Fig.~\ref{Case2}(b) at $t = 30$ s) and 0.07 with $\lambda = 0.3$ mm (Fig.~\ref{Case2}(c) at $t = 30$ s). While the absolute decrement in the mixing index appears minimal due to the system's inherently low baseline MI of 0.13, the primary utility of the acoustic field in this Configuration is not merely mixing suppression. Rather, its significance lies in the capacity to inhibit the gravity-driven transition toward equilibrium and stabilize the fluids in their initial vertically stratified state.
\begin{figure}[h!]
\centering
\includegraphics[width=1\linewidth]{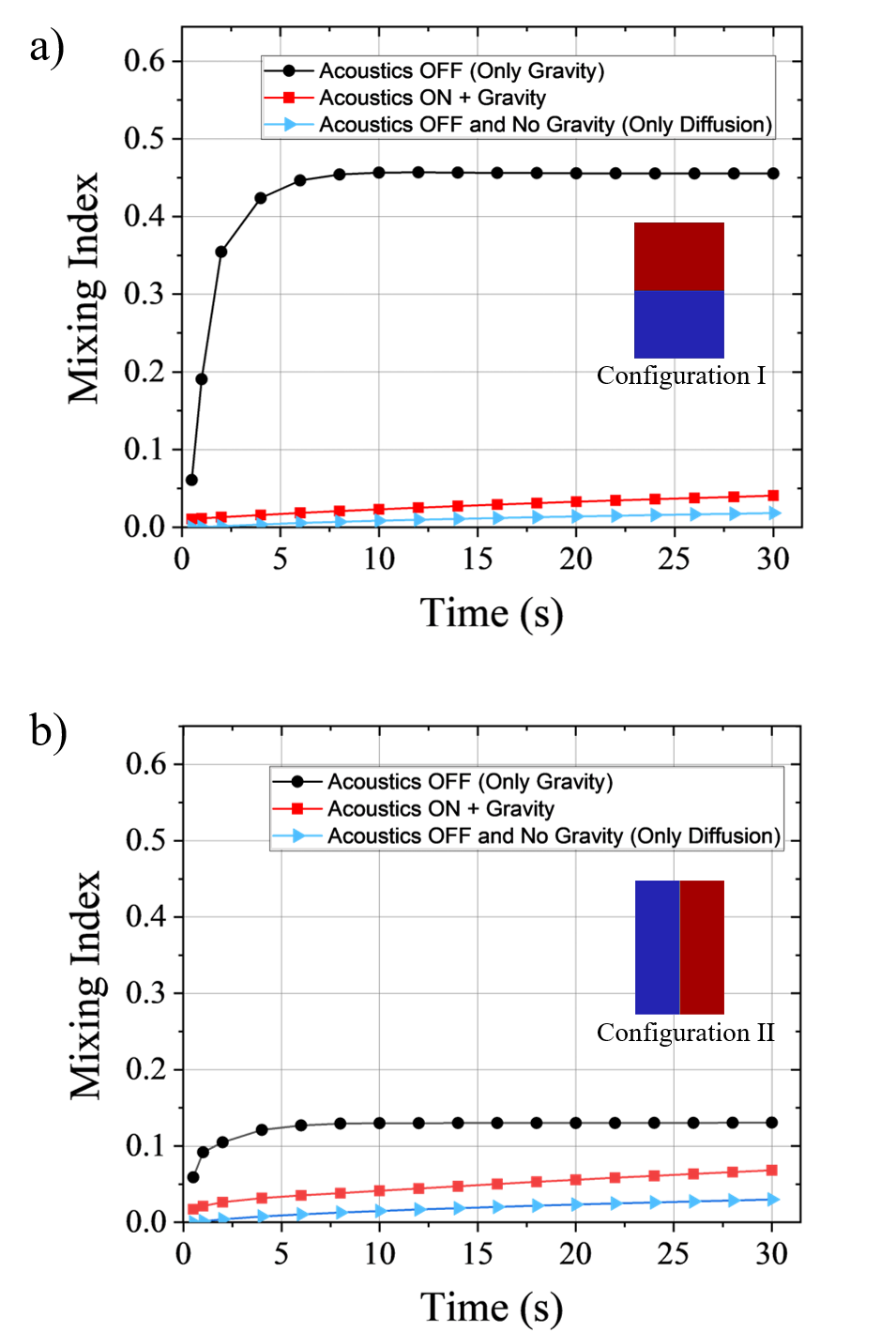}
\caption{Time evolution of the mixing index (MI) for (a) Configuration I and (b) Configuration II. The blue points correspond to the idealized baseline mixing condition, where mixing is only due to diffusion ($E_{ac}=0$ and $g=0$). The black points indicate the baseline mixing condition ($E_{ac}=0$), where mixing is induced solely by gravity. The red points represent the acoustically dominated regime ($E_{ac} > E_{cr}$) in the presence of gravity, where standing bulk acoustic waves ($E_{ac} = 125$ J/m³, $\lambda = 0.3$ mm) are applied perpendicular to the fluid interface.}
\label{Suppresion1}
\end{figure}

Below the critical energy density ($E_{ac} < E_{cr}$), the system enters the sub-critical, gravity-dominated regime. Here, the weak acoustic field triggers the same counterintuitive system response as observed in Configuration I, actively enhancing fluid mixing rather than suppressing it. This enhancement drives an increase in the mixing index (MI) up to 0.41 (Fig.~\ref{Case2}(d) at $t = 30$ s) and 0.42 (Fig.~\ref{Case2}(e) at $t = 30$ s) for the 1 mm and 0.3 mm wavelengths, respectively. The temporal evolution of the mixing dynamics for Configurations I and II, discussed above, is illustrated in Figs.~\ref{Suppresion1}(a) and (b), respectively. These plots provide a direct quantitative comparison across the idealized baseline condition ($E_{ac} = 0$ and $g = 0$), the gravity-driven baseline condition ($E_{ac} = 0$), and the acoustically dominated regime ($E_{ac} > E_{cr}$) with standing waves oriented perpendicular to the fluid interface.

From the preceding sections, it is evident that the acoustic stabilization of stratified fluids requires applying standing waves perpendicular to the fluid-fluid interface, with an acoustic energy density above the critical threshold ($E_{ac} > E_{cr}$). The subsequent sections detail the effects of varying the acoustic energy density and wavelength ($\lambda$) on the mixing dynamics.

\subsection{The Effect of Acoustic Energy Density ($E_{ac}$)}\label{Eac analysis}

\begin{figure}
\centering
\includegraphics[width=1\linewidth]{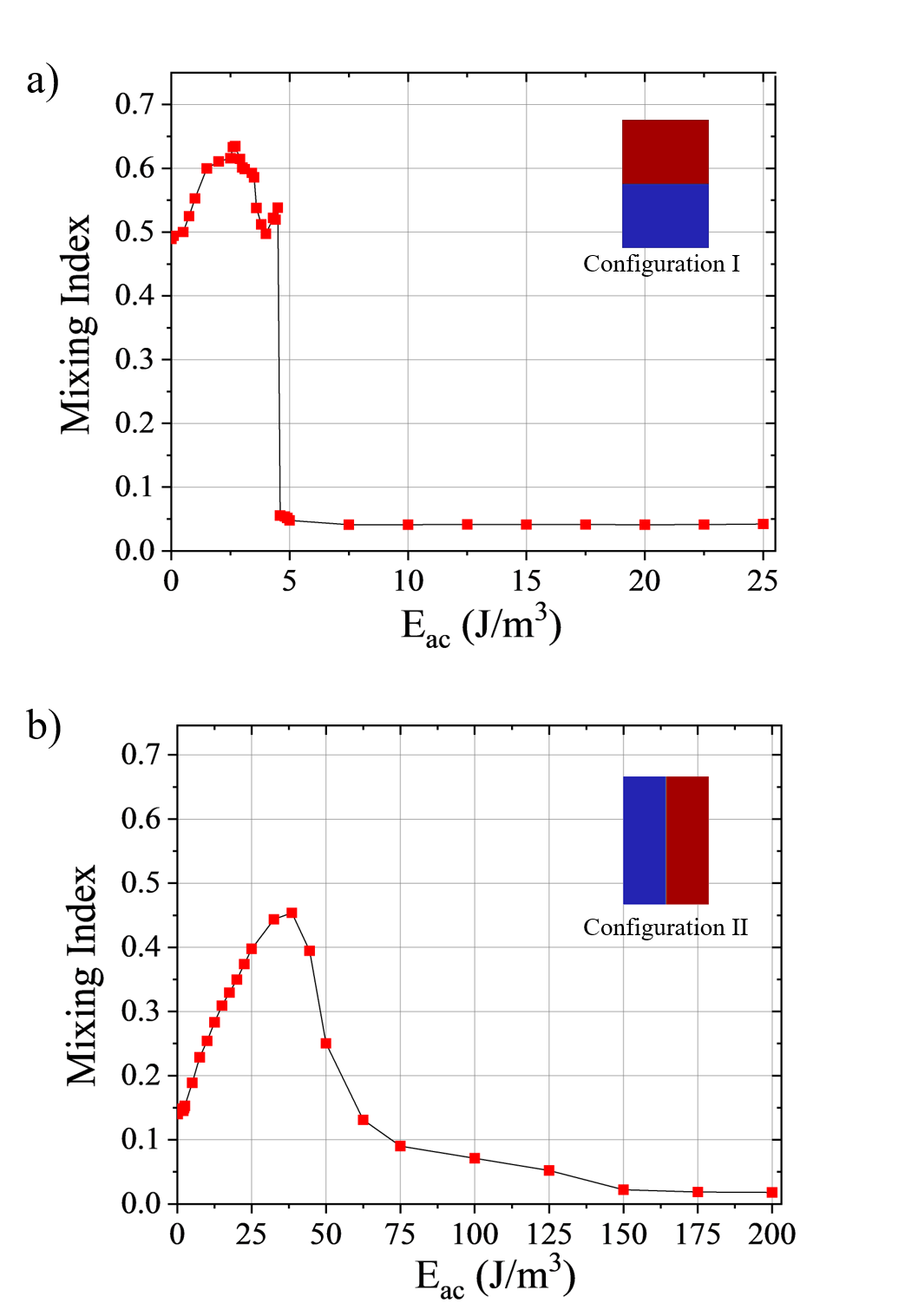}
\caption{Effect of acoustic energy density ($E_{ac}$) on fluid mixing in the presence of gravity, at an acoustic wavelength of 0.3 mm for (a) Configuration I and (b) Configuration II. The critical thresholds required to suppress fluid mixing are identified as $E_{cr} = 4.6$ J/m$^3$ for Configuration I and $E_{cr} = 62.5$ J/m$^3$ for Configuration II. Each data point represents the mixing index evaluated at $t = 30$ s for an applied $E_{ac}$.}
\label{case1&3_0.3mm}
\end{figure}

Figs.~\ref{case1&3_0.3mm}(a) and (b) illustrate the influence of acoustic energy density ($E_{ac}$) on fluid mixing for Configurations I and II, respectively. The results reveal the existence of a critical acoustic energy density ($E_{cr}$) and a distinct trend in the MI values (as shown in Fig.~\ref{case1&3_0.3mm}). The mixing induced solely by gravity establishes the baseline mixing condition ($E_{ac} = 0$, Figs.~\ref{Case1}(a) and \ref{Case2}(a)). The introduction of acoustic waves with $E_{ac} < E_{cr}$ (gravity-dominated regime) initially amplifies convective mixing, causing the MI to rise to a peak (as shown in Figs.~\ref{case1&3_0.3mm}(a) and (b)). Following this peak, the MI declines but remains higher than the baseline condition (as shown in Figs.~\ref{case1&3_0.3mm}(a) and (b)). As the applied $E_{ac}$ increases and eventually reaches the critical threshold ($E_{cr}$), the system transitions into the acoustically dominated regime ($E_{ac} > E_{cr}$). Under this condition ($E_{ac} > E_{cr}$), acoustic forces overcome the destabilizing gravitational forces, inhibiting the Rayleigh-Taylor instability and suppressing gravity-induced fluid mixing (Figs.~\ref{case1&3_0.3mm}(a) and (b)).

In the absence of an acoustic field ($E_{ac}=0$), fluid mixing in Configuration II is significantly lower than in Configuration I (MI = 0.13 and 0.45, respectively; Figs.~\ref{case1&3_0.3mm}(b) and (a)). This variation stems from their differing equilibrium transitions. In Configuration I, Rayleigh-Taylor instability disrupts the initially unstable equilibrium, causing perpendicular gravitational convection (Fig.~\ref{Case1}(a) at $t=0.4$ s). This generates multiple fluid-fluid mixing layers and chaotic mixing (Fig.~\ref{Case1}(a) at $t=2$ s), yielding an MI of 0.45 at $t=30$ s. Conversely, the vertically stratified, non-equilibrium initial state of Configuration II induces diagonal bulk fluid flow (Fig.~\ref{Case2}(a) at $t=0.1$ s). This rapidly drives the fluids into a stable, horizontally stratified equilibrium (Fig.~\ref{Case2}(a) at $t=2$ s) with minimal mixing layers, yielding the lower MI of 0.13 at $t=30$ s.

Achieving interfacial acoustic pinning in Configuration II demands a substantially higher critical acoustic energy ($E_{cr}$) compared to Configuration I ($E_{cr} = 4.6 $ J/m$^3$ and $62.5$ J/m$^3$ for Configurations I and II, respectively, at $\lambda = 0.3$ mm; Figs.~\ref{case1&3_0.3mm}(b) and (a)). While Configuration I begins in an equilibrium state (even though unstable), its initial net gravitational driving force is minimal, thus requiring a significantly lower acoustic energy threshold. Whereas Configuration II initiates from a non-equilibrium state characterized by strong, unbalanced gravitational forces; hence, it demands a much larger acoustic energy threshold to successfully arrest the fluids in their initial non-equilibrium state. Furthermore, the value of $E_{cr}$ is directly dependent upon the channel Configuration, initial fluid positioning, channel aspect ratio, and the governing forces acting within the system. A systematic investigation of these parameters remains a subject for future research.

\subsection{\textbf{The Effect of Acoustic Wavelength ($\lambda$)}}\label{frequency analysis}

\begin{figure}[h!]
\centering
\includegraphics[width=1\linewidth]{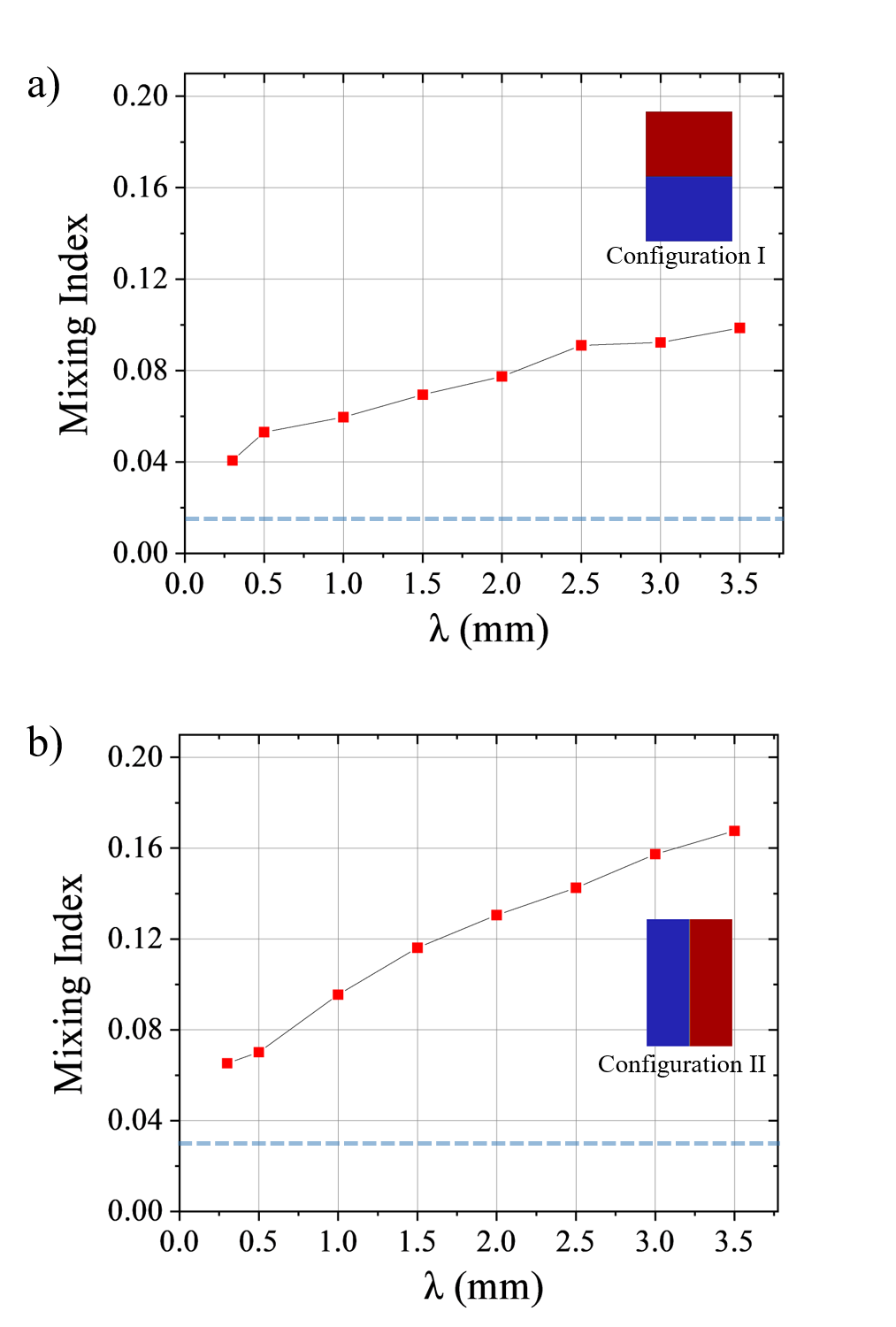}
\caption{Effect of acoustic wavelength on the fluid mixing in presence of gravity, at a constant acoustic energy density of $E_{ac} = 125$ J/m$^3$. Results are shown for (a) Configuration I and (b) Configuration II. The data points indicate the mixing index evaluated at each applied acoustic wavelength at $t = 30$ s. The blue dashed line denotes the idealized baseline mixing condition governed solely by diffusion ($E_{ac}=0$ and $g=0$), with values of 0.018 and 0.029 for Configurations I and II, respectively.}
\label{wave1}
\end{figure}

For the results presented in this section, the acoustic energy density is maintained at $E_{ac} = 125$ J/m$^3$, ensuring an acoustically dominated regime ($E_{ac} > E_{cr}$) for both Configurations I and II. Under this condition, Fig.\ref{wave1}(a) and (b) illustrate the influence of acoustic wavelength ($\lambda$) on fluid mixing for Configurations I and II, respectively. The acoustic wavelength $\lambda$ determines the distance between the interface and the nearest pressure node or antinode. For a larger wavelength, this distance increases, thereby causing a greater volume of fluid to relocate/convect to form the acoustically controlled fluid barrier. The resulting expansion of this barrier zone provides a larger volume of fluid available for mixing via diffusion. Consequently, as observed in both configurations, increasing the wavelength directly increases the mixing index (as shown in Fig.\ref{wave1}(a) and (b)).

The comparison of the results shown in Figs.~\ref{wave1}(a) and (b) under the $E_{ac} > E_{cr}$ condition reveals that Configuration II consistently yields a higher mixing index than Configuration I across all evaluated wavelengths. This pattern is observed because the rectangular channel selected for this study has a height greater than its width ($H=6$ mm, $W=4$ mm). Consequently, the total volume of fluid mixed during the formation of the acoustically controlled fluid barrier is higher for Configuration II (Figs.\ref{wave1}(a) and (b)). The acoustic field drives the relocation of fluid layers along the longer vertical height for Configuration II, (Fig.~\ref{Case2}(b), $t = 0.5$~s), creating a larger interfacial contact area for molecular diffusion. In contrast, acoustic relocation in Configuration I occurs across the shorter horizontal width, resulting in a smaller contact area and reduced fluid mixing relative to Configuration II (Fig.~\ref{Case1}(b), $t = 0.15$~s).

The idealized baseline condition (mixing solely due to molecular diffusion; $E_{ac}=0$ and $g=0$), represented by the blue lines in Figs.~\ref{wave1}(a) and (b), establishes the theoretical minimum achievable fluid mixing for Configurations I and II, with corresponding MI values of 0.018 and 0.029, respectively. Irrespective of the applied acoustic energy density ($E_{ac}$) or the selected acoustic wavelength ($\lambda$), the mixing index cannot be reduced beyond the fundamental lower bound achieved under this idealized baseline condition ($E_{ac}=0$ and $g=0$). While examining the variation in acoustic wavelength ($\lambda$), it is observed that for Configuration I, extending the wavelength even to 3.5~mm does not cause the fluid mixing (Fig.~\ref{wave1}(a), MI = 0.1 at $t = 30$~s) to be greater than the mixing observed under its baseline condition (Fig.~\ref{Case1}(a), MI = 0.45). This behavior is due to the chaotic mixing inherent to this baseline condition, which yields a highly elevated MI of 0.45 at $t = 30$~s (Fig.~\ref{Case1}(a)). Conversely, Configuration II exhibits a lower baseline MI of 0.13 at $t = 30$~s (Fig.~\ref{Case2}(a)), which is exceeded at higher wavelengths by mixing between the larger acoustically relocated fluid layers (Fig.~\ref{wave1}(b)).

\section{\textbf{Conclusion}}\label{sec4} 

This theoretical study has established the application of the acoustic body force as a robust mechanism for stabilizing stratified miscible fluids against gravity in minichannels. By analyzing the interplay between gravitational and acoustic forces, we demonstrated that achieving this acoustofluidic suppression of fluid mixing directly depends on the direction of wave application, the acoustic energy density ($E_{ac}$), and the acoustic wavelength ($\lambda$). The findings of this study present a method for the active control of gravitational instabilities in systems where buoyancy-driven mixing must be restricted for targeted durations. Specifically, in Configuration I, the acoustic field efficiently suppressed gravity-driven mixing, reducing the mixing index (MI) from 0.45 to 0.04. This suppression mechanism can be utilized in the development of dynamic concentration gradient generators and highly customized minifluidic lab-on-a-chip platforms. Future research extending from this work has scope in two main directions. First, subsequent investigations can focus on the experimental demonstration of the predicted stabilization dynamics. Second, the present work assumes that variations in fluid properties on a slow timescale do not influence the acoustic field. However, in general, fluid dynamics and acoustics are bidirectionally coupled. Further analysis is required to incorporate this bidirectional coupling and explore its influence on the acoustic stabilization mechanism. Additionally, this theoretical study can be extended to investigate the influence of channel geometries and aspect ratios on acoustofluidic mixing suppression.

\section*{Author Declarations}
\subsection*{Conflict of Interest}
The authors have no conflicts of interest to disclose.

\subsection*{Acknowledgment} This work is supported by the Department of Science \& Technology - Fund for Improvement of Science \& Technology Infrastructure (DST-FIST) via Grant No: SR/FST/ET-I/2021/815.

$^\dagger$ These authors have contributed equally to this work.

\subsection*{DATA AVAILABILITY}
The data that support the findings of this study are available from
the corresponding author upon reasonable request.

\appendix
\section{FLUID PROPERTIES}
\label{appendix:A}
\begin{table}[H]
\centering
\renewcommand{\arraystretch}{1.05}
\setlength{\tabcolsep}{3pt}
\scriptsize
\begin{tabular}{|c|c|}
\hline
Fluid properties & Expression in terms of the concentration of Iodixanol ($\phi$)\\
\hline
Density ($\rho_0$) & $(5.245 \times \phi) + 1005$ \\
(kg/m$^3$) & \\
\hline
Viscosity ($\mu$) & $(2.173 \times 10^{-8} \times \phi^3) + (2.419 \times 10^{-7} \times \phi^2) + (1.952 \times 10^{-5} \times \phi)$\\
(Pa-s) & + $(9.540 \times 10^{-4})$\\
\hline
Speed of sound $(c_0)$ & $(2.557 \times 10^{-5} \times \phi^3) + (8.053 \times 10^{-3} \times \phi^2)$\\
(m/s) & - $(0.7308 \times \phi) + 1507$\\
\hline
\end{tabular}
\caption{Fluid properties in terms of iodixanol concentration $\phi$ (Augustsson, P. et al.~\cite{AugustssonKarlsenSuBruusVoldman2016}).}
\label{tab:1}
\end{table}

The equations presented in Table~\ref{tab:1} were directly integrated into the computational model, enabling the dynamic calculation of these fluid properties throughout the simulation. In this setup, the selected fluids exhibit a density difference ($\Delta \rho \approx 100$~kg/m$^3$) which is larger than their speed of sound difference ($\Delta c_0 \approx 8$~m/s). The diffusion coefficient for iodixanol is defined as $D = 2.5 \times 10^{-10}\ \mathrm{m^2/s}$.\cite{NairKimBraatzStrano2008}

\section*{References}
\nocite{*}

\bibliography{References}

\end{document}